\documentclass[10pt,journal,compsoc, onecolumn]{IEEEtran}
 \pdfoutput=1
\usepackage{graphicx}
\usepackage[utf8]{inputenc}

\usepackage[procnames]{listings}
\usepackage{color}

\usepackage{amssymb}
\graphicspath{{figures/}}

\usepackage{url}
\usepackage{wrapfig}
\usepackage{units}

\usepackage[printonlyused]{acronym}

\usepackage{calc}
\usepackage{amsmath}
\usepackage{amssymb}
\usepackage{amsfonts}

\usepackage{algorithm}
\usepackage{algpseudocode}
\usepackage{color}
\usepackage[table]{xcolor} 
\usepackage{caption}
\usepackage{subcaption}
\usepackage{stfloats}
\usepackage{nicefrac}

\usepackage{booktabs}
\newcolumntype{C}[1]{>{\centering\arraybackslash}p{#1}}

\usepackage{color}

\newcolumntype{C}[1]{>{\centering\arraybackslash}p{#1}}

\usepackage{booktabs}

\definecolor{BgGray}{gray}{0.7}%
\definecolor{BgGray2}{gray}{0.96}%
\definecolor{RowColorOdd}{named}{BgGray2}%
\definecolor{RowColorEven}{named}{white}%
\definecolor{comments}{gray}{.5}
\definecolor{Gray}{gray}{0.85}

\definecolor{keywords}{RGB}{255,0,90}
\definecolor{red}{RGB}{160,0,0}
\definecolor{green}{RGB}{0,150,0}
\definecolor{deepblue}{rgb}{0,0,0.5}
\definecolor{deepred}{rgb}{0.6,0,0}
\definecolor{deepgreen}{rgb}{0,0.5,0}

\usepackage{multirow}

\usepackage[utf8]{inputenc}
\usepackage{tabularx}
\usepackage{multicol}
\usepackage{graphicx}

\usepackage{tikz}

\usepackage[headsepline,plainheadsepline,footsepline,plainfootsepline]{scrpage2}
\usepackage{mathptmx}
\usepackage[scaled=.92]{helvet}
\usepackage{courier}
\usepackage[multiuser,nomargin,marginclue,footnote]{fixme}

\definecolor{BgGray}{gray}{0.1}%
\definecolor{BgGray2}{gray}{0.96}%
\definecolor{RowColorOdd}{named}{BgGray2}%
\definecolor{RowColorEven}{named}{white}%
\definecolor{comments}{gray}{.5}
\definecolor{Gray}{gray}{0.85}

\usepackage{pifont}
\usepackage[multiuser,nomargin,marginclue,footnote]{fixme}
\fxusetheme{color}
\fxsetup{targetlayout=colorcb}
\FXRegisterAuthor{all}{anall}{ALL}
\FXRegisterAuthor{tolja}{antolja}{TOLJA}
\FXRegisterAuthor{mch}{anmc}{MC}
\FXRegisterAuthor{cp}{ancp}{CP}

\fxsetup{status=draft}

\definecolor{BgGray}{gray}{0.5}%

\usepackage{geometry}
\clearscrheadfoot

\ihead[\Large {\scshape TKN Technical Reports Series / Technische Universität Berlin }]{\small {TR \trnumber}}

\ifoot[{\tiny \begin{minipage}{4.0cm}Copyright at Technische Universität Berlin.\newline All Rights Reserved.\end{minipage}}]{{\tiny \begin{minipage}{4.0cm}Copyright at Technische Universität Berlin.\newline All Rights Reserved.\end{minipage}}}
\ofoot[Page \pagemark]{Page \pagemark}

\newcommand{\trnumber}{TKN-16-0004}
\newcommand{\trdate}{November 2016}
\newcommand{\trauthor}{Sven Zehl, Anatolij Zubow and Adam Wolisz}
\newcommand{\tremail}{\{zehl, zubow, wolisz\}@tkn.tu-berlin.de}
\newcommand{\trtitle}{hMAC: Enabling Hybrid TDMA/CSMA on IEEE 802.11 Hardware}

\ifCLASSOPTIONcompsoc 
  \usepackage[nocompress]{cite}
\else
  \usepackage{cite}
\fi

\ifCLASSINFOpdf
\else
\fi

\begin{document}




{
\sffamily

\thispagestyle{empty}

\begin{tabularx}{\columnwidth}{cXc}
  \includegraphics[height=1cm]{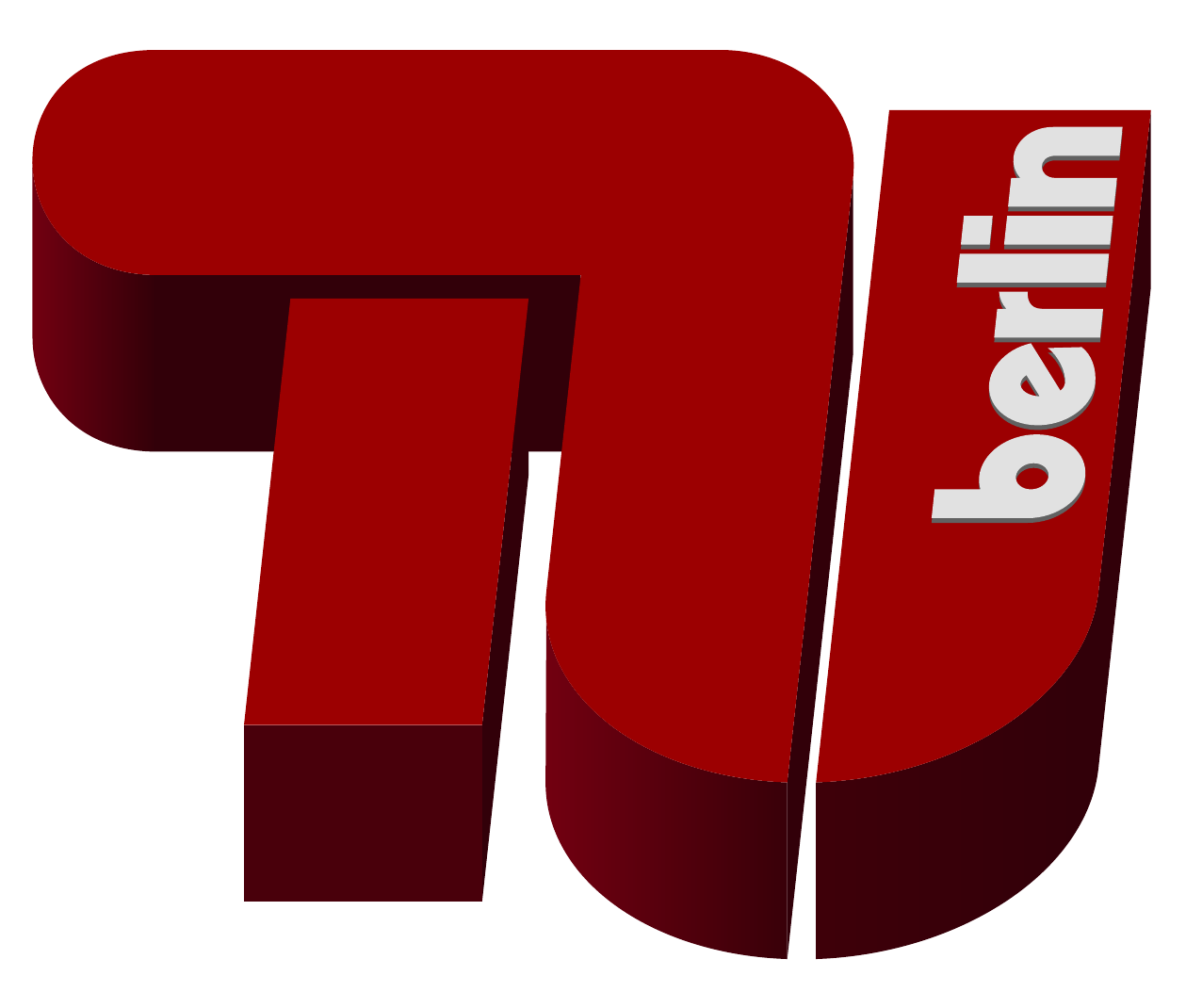}
  & &
  \includegraphics[height=1cm]{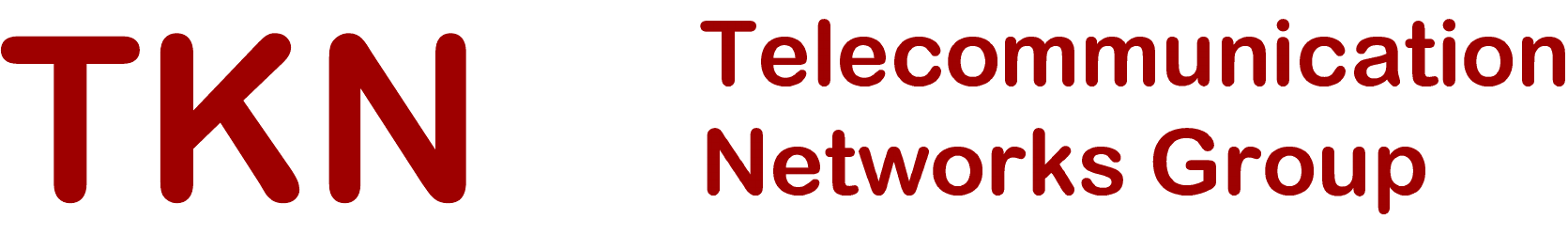}
  \\
\end{tabularx}

\vspace{1.0cm}

\begin{center}
{\huge
\noindent
Technische Universität Berlin

\vspace{0.5cm}

\noindent
Telecommunication Networks Group

\begin{center}
\rule{15.5cm}{0.4pt}
\end{center}
}
\end{center}

\begin{minipage}[][11.0cm][c]{14.5cm}
{\Huge

\begin{center}
\trtitle
\end{center}

\begin{center}
{\LARGE \trauthor} \\
{\Large \tremail}
\end{center}

\begin{center}
Berlin, \trdate
\end{center}

\vspace{0.5cm}

}

\begin{center}
\setlength{\fboxrule}{2pt}\setlength{\fboxsep}{2mm}
\fbox{TKN Technical Report \trnumber}
\end{center}

\end{minipage}

\setlength{\fboxrule}{0.4pt}
\setlength{\fboxsep}{0.4pt}

\begin{center}

  \rule{15.5cm}{0.4pt}

  \vspace{0.5cm}

  {\huge {TKN Technical Reports Series}}

  \vspace{0.5cm}

  {\huge Editor: Prof. Dr.-Ing. Adam Wolisz}

  \vspace{0.5cm}

 \end{center}

}

\title{}
\author{}

\newpage
\begin{abstract}
\normalsize 
We present our current work-in-progress on the design and implementation of a hybrid TDMA/CSMA medium access architecture, hereafter referred to as hMAC, which can be used on top of commercial IEEE 802.11 off-the-shelf hardware. The software only solution is based on the popular Linux ATH9K softMAC driver and hence can be used with standard Linux systems using Atheros based wireless network devices. 

The proposed hMAC exploits the standard 802.11 power saving functionality present in the ATH9K device driver to enable control of the software packet queues. This allows the assignment of TDMA time slots on wireless link and traffic class basis. While the solution is placed only in the device driver, the CSMA/CA functionality on hardware level is still active. This enables inter-working with standard unmodified 802.11 devices. 

We tested our prototypical hMAC implementation in a small test-bed. Therefore, we implemented a centralized interference management scheme in which pairs of links suffering from a hidden node problem are assigned to TDMA time slots on a per-link basis. To show the benefits of the proposed hMAC approach we compared the results with standard 802.11 DCF and classical, i.e. per-node, TDMA. Finally, to enable collaboration with the research community, the hMAC source code is provided as open-source.
\end{abstract}

\begin{IEEEkeywords}
\center WiFi, IEEE 802.11, TDMA, CSMA/CA, Interference Management, Slicing, COTS, ATH9k, Hidden Node Mitigation, per-link-TDMA
\end{IEEEkeywords}


\newpage
%
%
\section{Introduction}\label{sec:introduction}

\begin{figure}[!ht]
   \begin{center}
       \includegraphics[width=\linewidth]{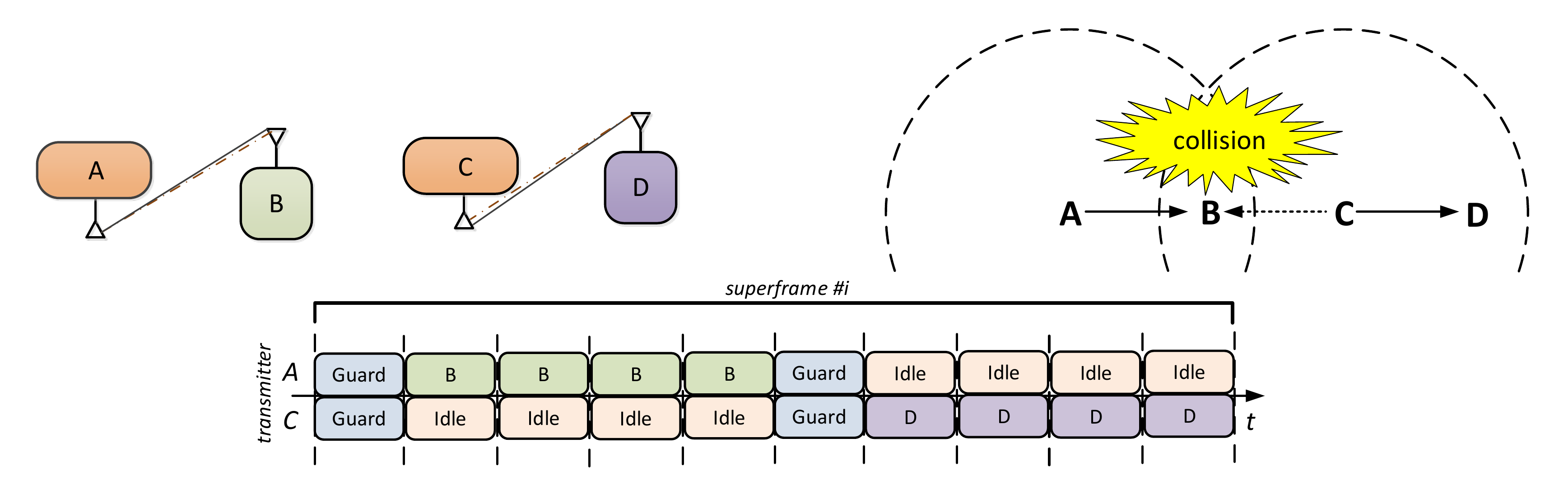}
   \end{center}
    \vspace{-10pt}
   \caption{Illustration of hidden node problem and interference avoidance through airtime management, i.e. assignment of exclusive time slots.}
   \label{fig:airtime_exp_simple}
\end{figure}

A widely known problem experienced in IEEE 802.11 (Wi-Fi) networks is performance degradation due to \textbf{co-channel interference} because of \textbf{hidden} nodes. The impact can be mitigated by preventing overlapping transmissions (in time) between co-located APs by airtime management through interference avoidance techniques, cf. Fig.~\ref{fig:airtime_exp_simple}.

Although the assignment of exclusive time slots to downlink (DL) traffic for AP1 and AP2 would solve the hidden node problem it is however \textbf{inefficient} as it unnecessarily prevents spatial reuse. This is because Access Points (AP) of a Wi-Fi infrastructure network usually do not serve single but rather multiple STAs. Simply putting the whole APs DL traffic separated creates unfairness to STAs not suffering from hidden node problem. 

\begin{figure}[!ht]
	\begin{center}
		\includegraphics[width=0.6\linewidth]{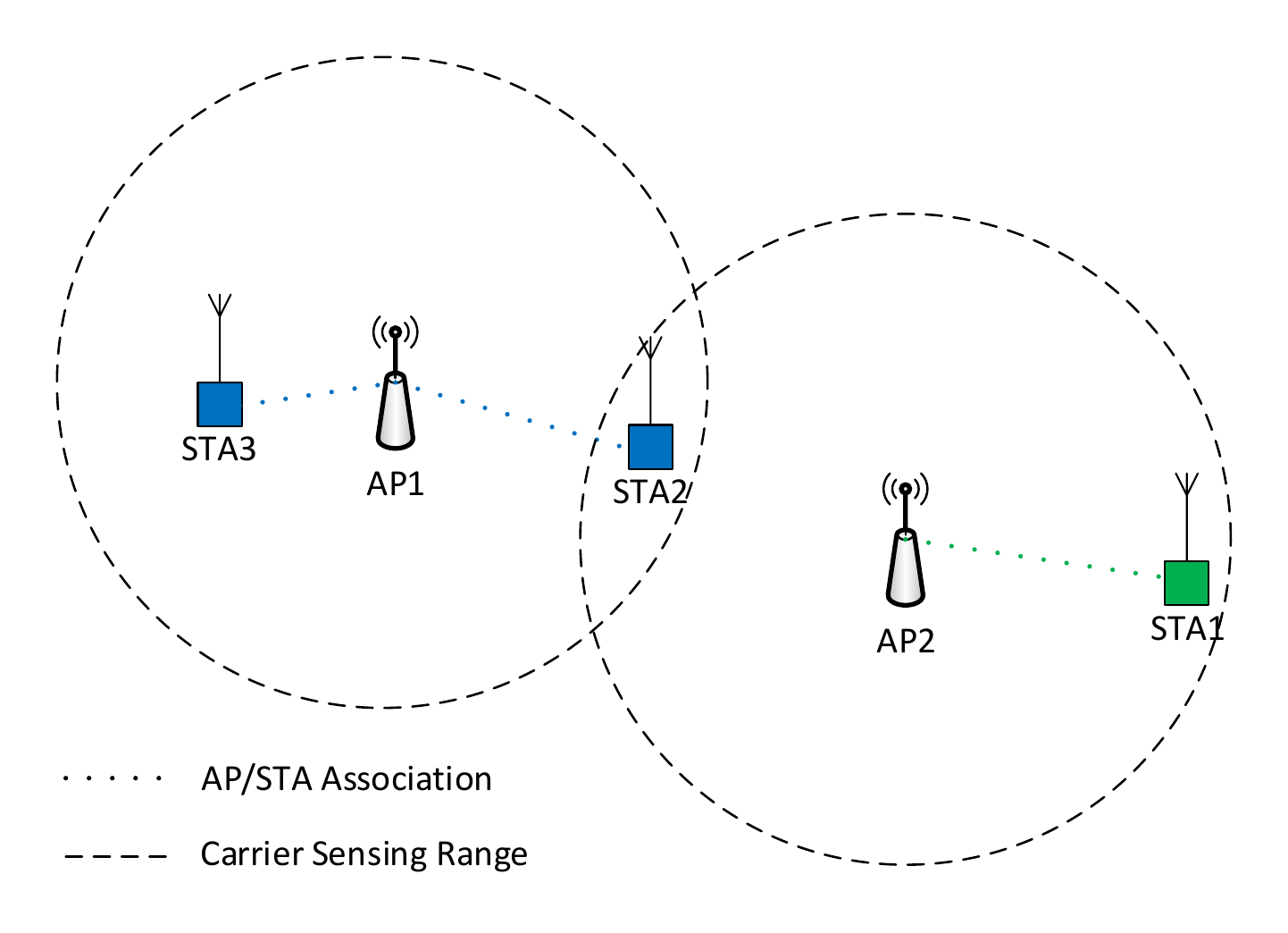}
	\end{center}
	\vspace{-10pt}
	\caption{\textbf{Example 1} - AP1 and AP2 are hidden to each other, i.e. collision at STA2 (only considering downlink traffic). In addition, AP2 and STA2 are in the exposed node scenario if uplink traffic of STAs is considered.}
	\label{fig:hidden-node}
\end{figure}

\begin{figure}
\centering
\begin{subfigure}{.5\textwidth}
  \centering
  \includegraphics[width=0.8\linewidth]{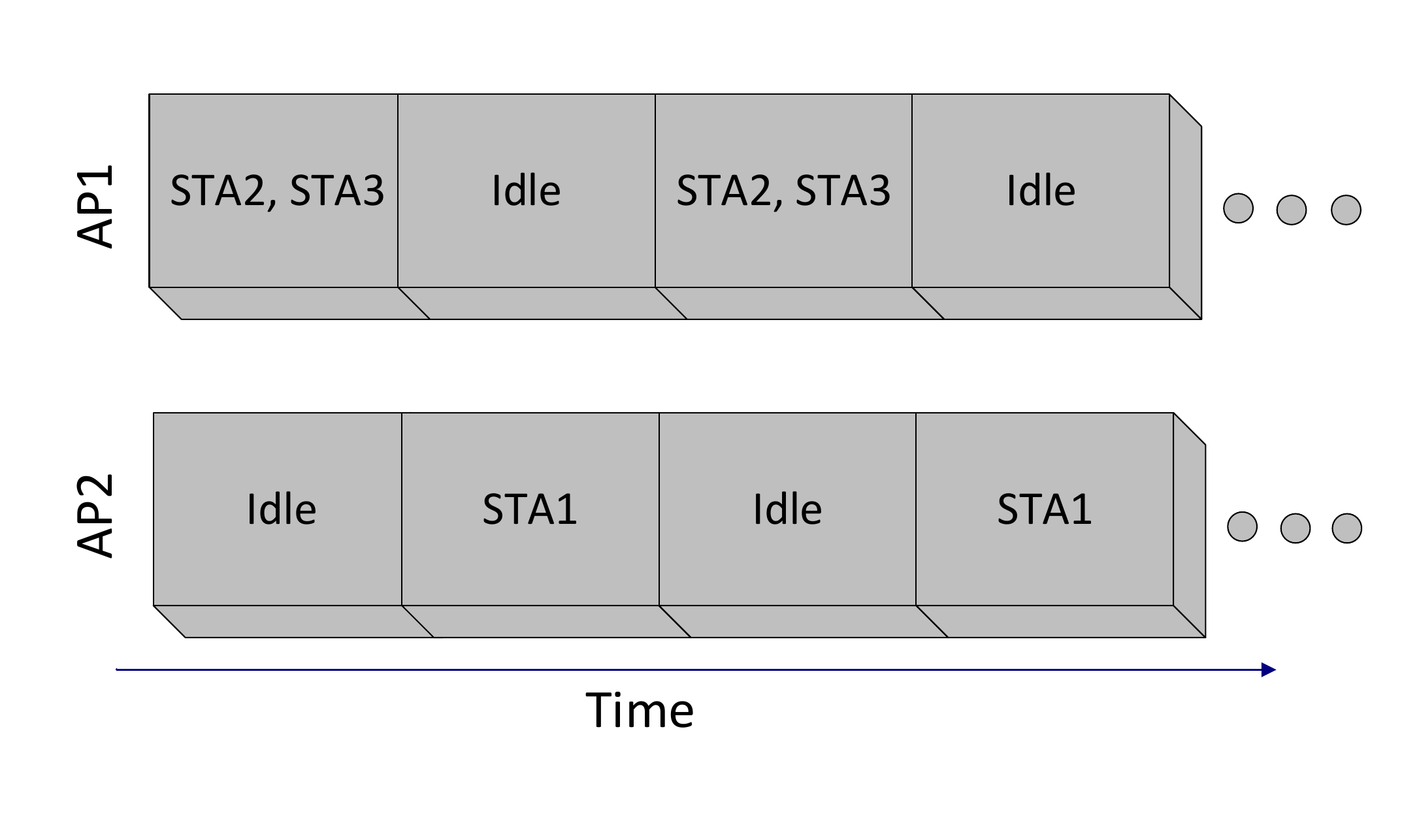}
  \caption{Classical TDMA - per node slot assignment.}
  \label{fig:classical_tdma}
\end{subfigure}%
\begin{subfigure}{.5\textwidth}
  \centering
  \includegraphics[width=0.8\linewidth]{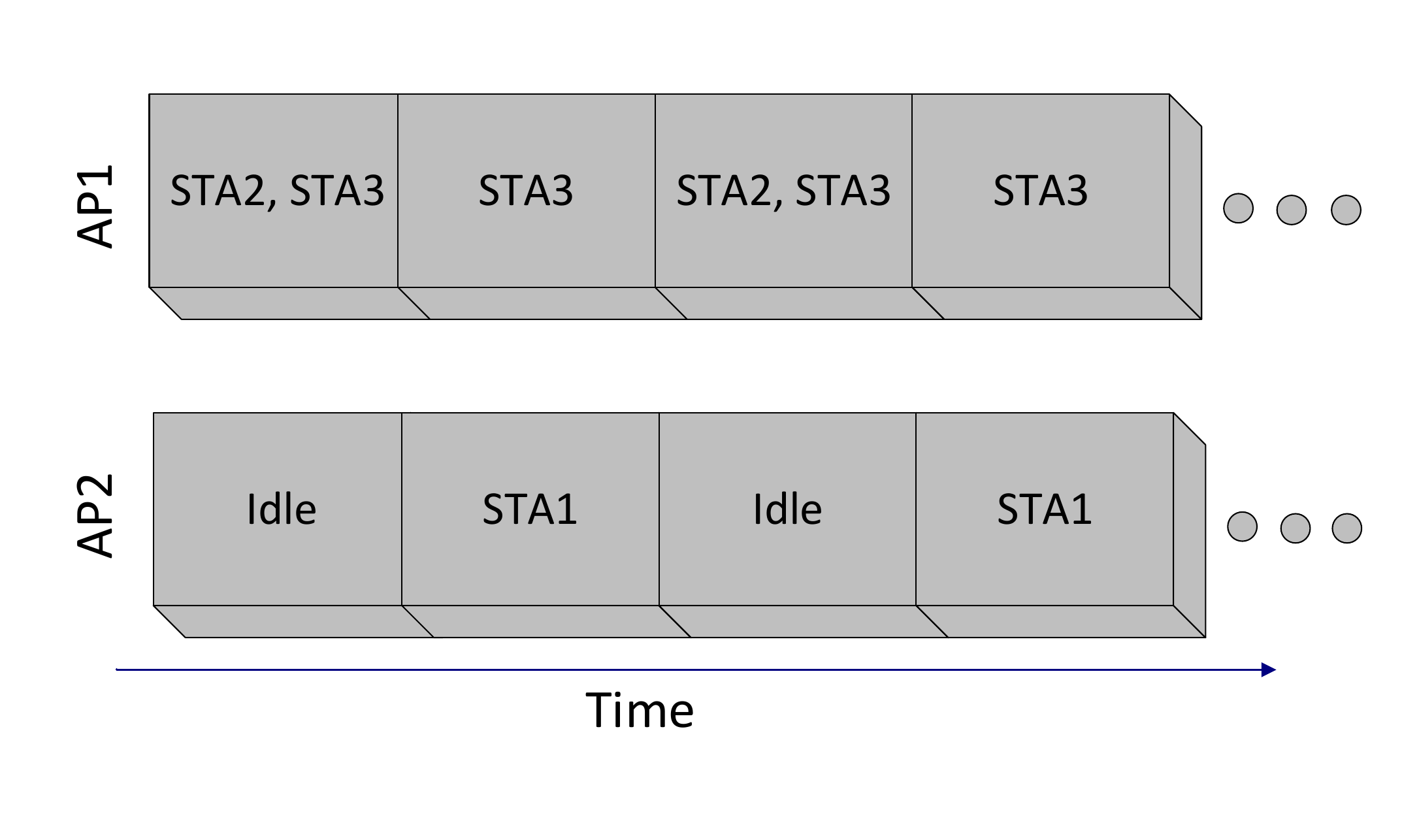}
  \caption{Proposed hMAC - per link slot assignment.}
  \label{fig:per_link_tdma_sched}
\end{subfigure}
\caption{Example TDMA schedules for scenario shown in Fig.~\ref{fig:hidden-node}.}
\label{fig:example_schedules1}
\end{figure}

The scenario shown in Fig.~\ref{fig:hidden-node} illustrates such a case. AP1 and AP2 are hidden to each other. If AP1 wants to transmit DL traffic to STA2 while AP2 transmits DL traffic to STA1, the frames of AP1 and AP2 will collide at STA2 \footnote{Moreover, if also the uplink traffic of the STAs is considered, AP2 and STA2 are in the exposed terminal case, in which they prevent each other from sending. Nevertheless, in this paper we focus on solving the hidden node problem and therefore only consider DL traffic from APs to STAs.}.
According to the classical approach as shown in Fig,~\ref{fig:classical_tdma}, the hidden node problem can be solved by separating the two APs in time domain, which means that AP1 can serve its two STAs, STA2 and STA3, only in every second time slot. Obviously, this is in inefficient as sending DL traffic from AP1 to STA3 does not create any significant interference to the link AP2-STA1 and therefore can be served concurrently resulting in increased spatial reuse.

As a consequence, we argue that a much \textbf{finer control} of TDMA is required. More specifically, instead of assigning TDMA slots on a per node basis (e.g. AP) we demand the possibility to assign \textbf{time slots} for each wireless link independently. For our scenario shown in Fig.~\ref{fig:hidden-node},  Fig.~\ref{fig:classical_tdma} and \ref{fig:per_link_tdma_sched} show example schedules when either TDMA configuration on a per-node basis (Fig.\ref{fig:classical_tdma}) or on the proposed per-link basis (Fig.\ref{fig:per_link_tdma_sched}) is applied.

\begin{figure}[!ht]
	\begin{center}
		\includegraphics[width=1\linewidth]{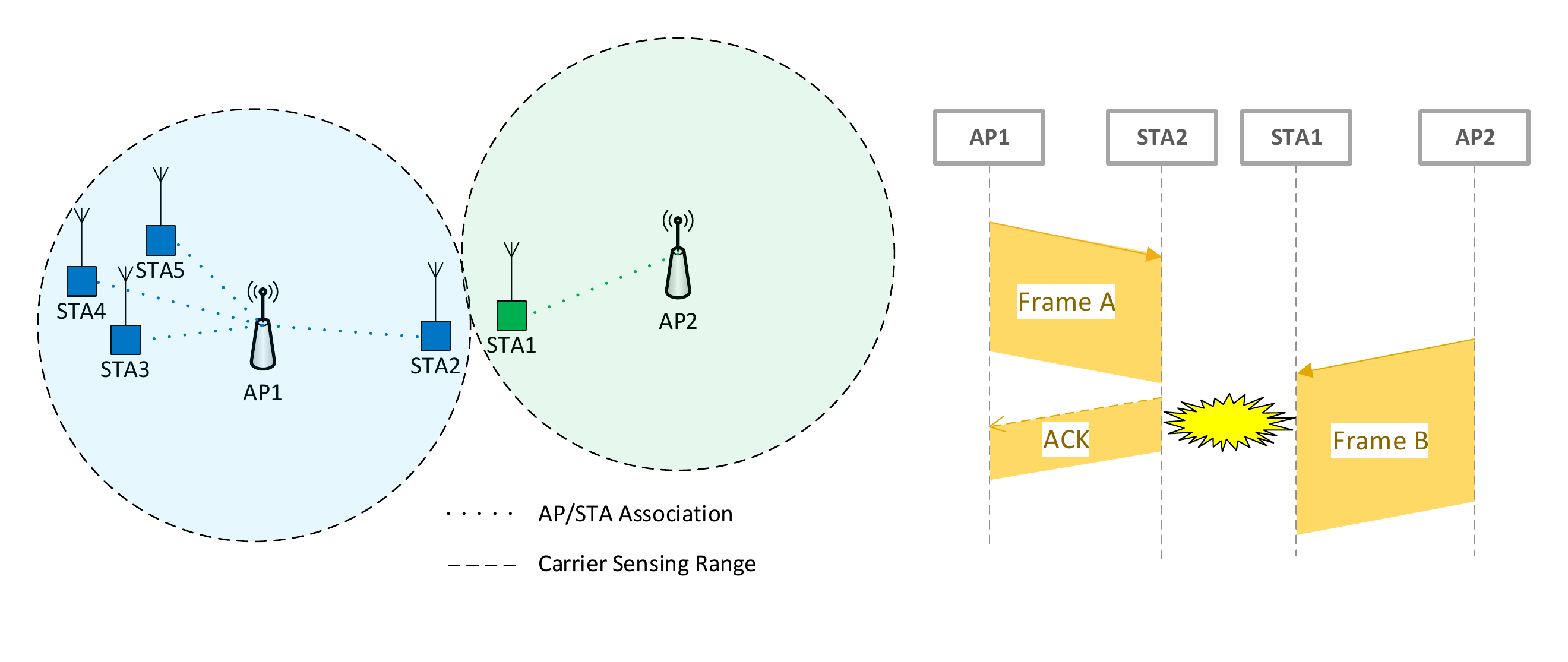}
	\end{center}
	\vspace{-10pt}
	\caption{\textbf{Example 2} - AP1 and AP2 are not hidden to each other, but ACK frames may collide with data frames (only considering downlink traffic).}
	\label{fig:no-hidden-node}
\end{figure}

Fig.~\ref{fig:no-hidden-node} illustrates another scenario where the assignment of TDMA slots on a per-link basis is more promising. As shown, even in a scenario in which two transmitter nodes are not hidden to each other, frame collisions can still occur. As the 802.11 protocol demands the acknowledgment of each unicast frame those ACK frames can collide with data packets as ACK frames are sent contention-free. In order to avoid degradation of network performance the packet transmissions on the two links, AP1-STA2 and AP2-STA1, need to be orthogonalized in time.

\begin{figure}
\centering
\begin{subfigure}{.5\textwidth}
  \centering
  \includegraphics[width=0.8\linewidth]{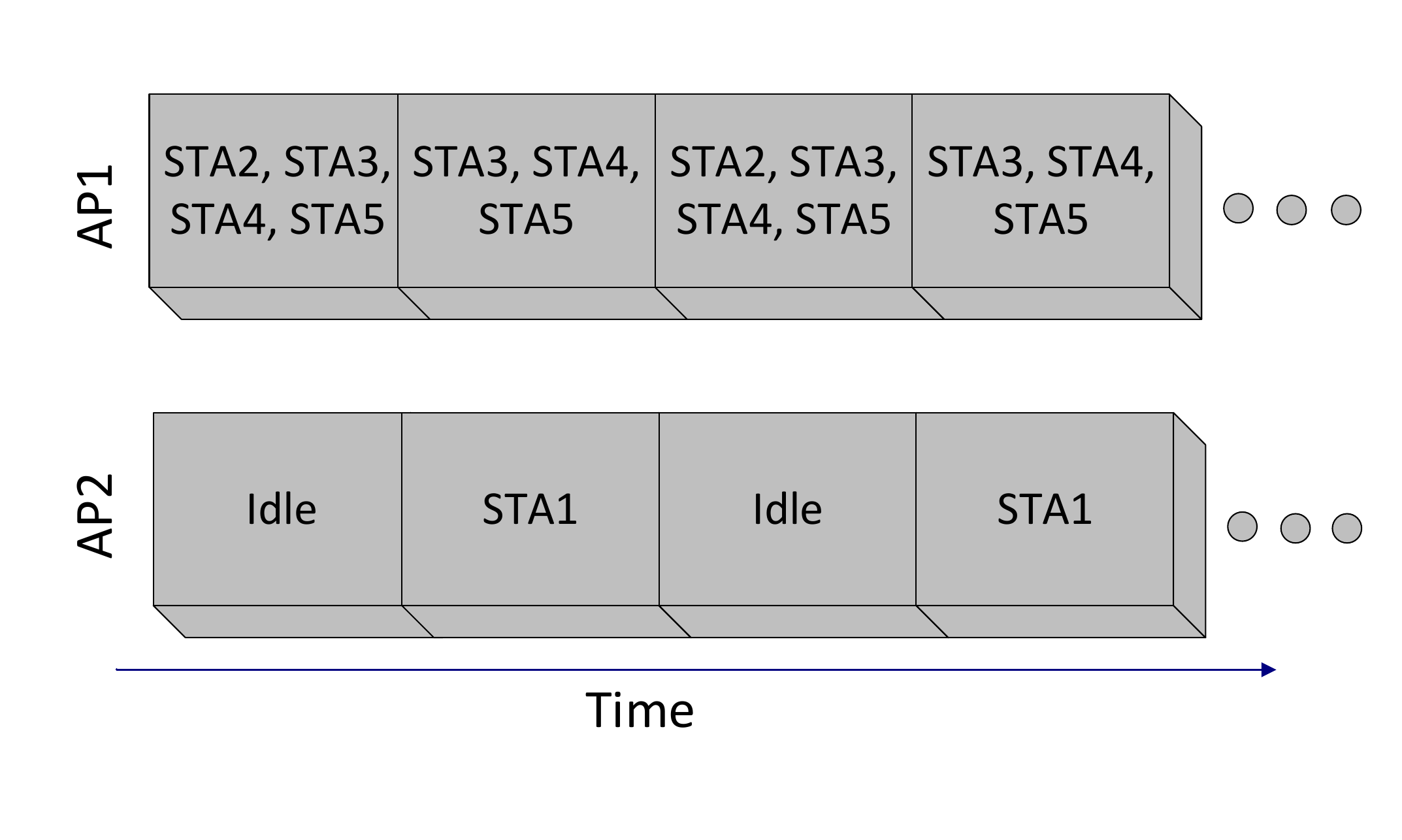}
  \caption{Per link TDMA schedule for scenario from Fig.~\ref{fig:no-hidden-node}.}
  \label{fig:link-tdma}
\end{subfigure}%
\begin{subfigure}{.5\textwidth}
  \centering
  \includegraphics[width=0.8\linewidth]{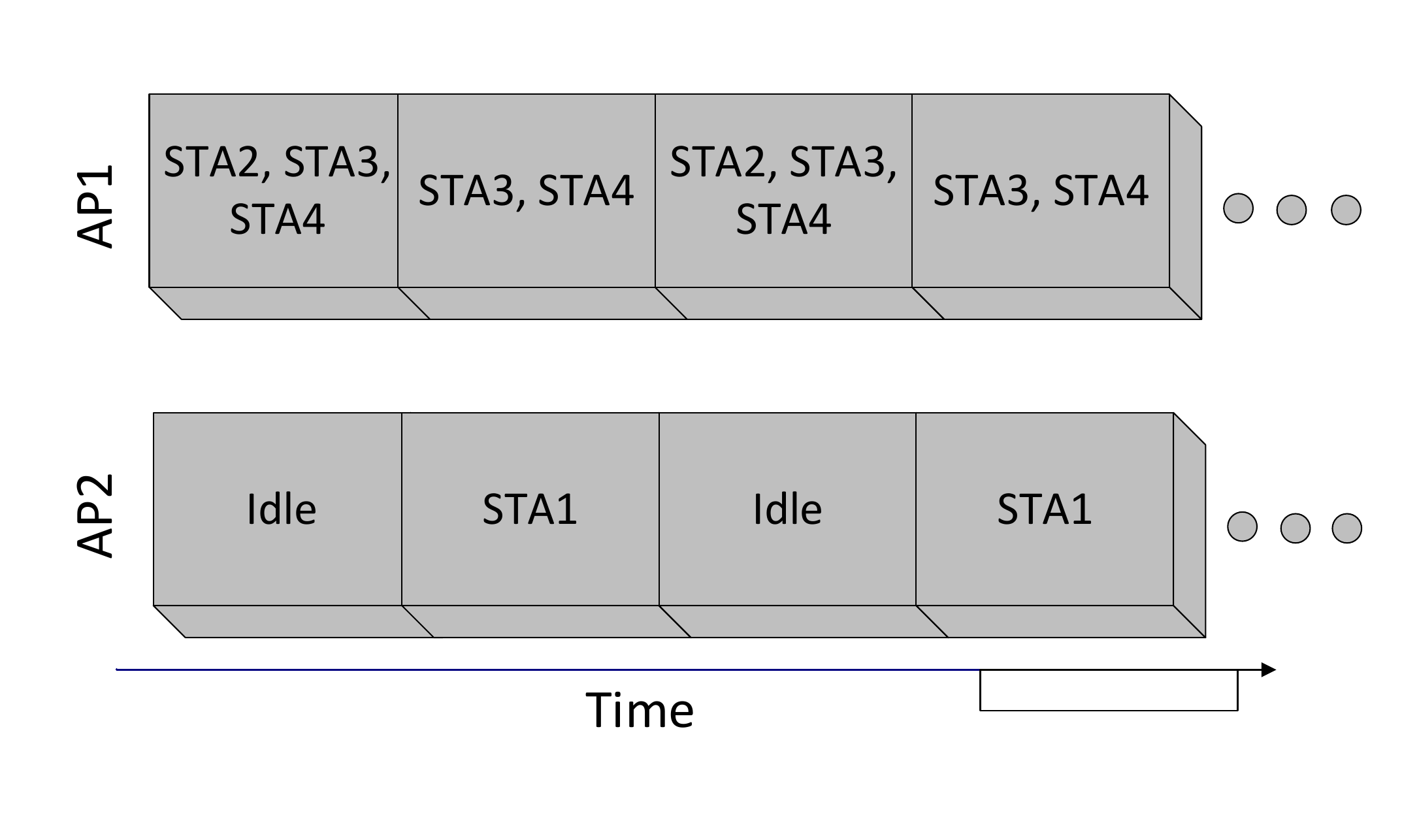}
  \caption{Per link TDMA schedule for scenario from Fig.~\ref{fig:link-tpc}}
  \label{fig:link-tdmab}
\end{subfigure}
\caption{Example TDMA schedules for scenarios shown in Fig.~\ref{fig:no-hidden-node} and Fig.~\ref{fig:link-tpc}.}
\label{fig:example_schedules}
\end{figure}

Again, also in this scenario, the assignment of exclusive TDMA time slots on a per links basis helps to mitigate frame collisions while not unnecessary preventing spatial reuse. Specifically, we allow the DL traffic towards STAs 3-4 to always take place while only transmissions on link AP1-STA2 are separated in time with the link AP2-STA1, e.g. Fig.~\ref{fig:link-tdma}.

\begin{figure}[!ht]
	\begin{center}
		\includegraphics[width=0.6\linewidth]{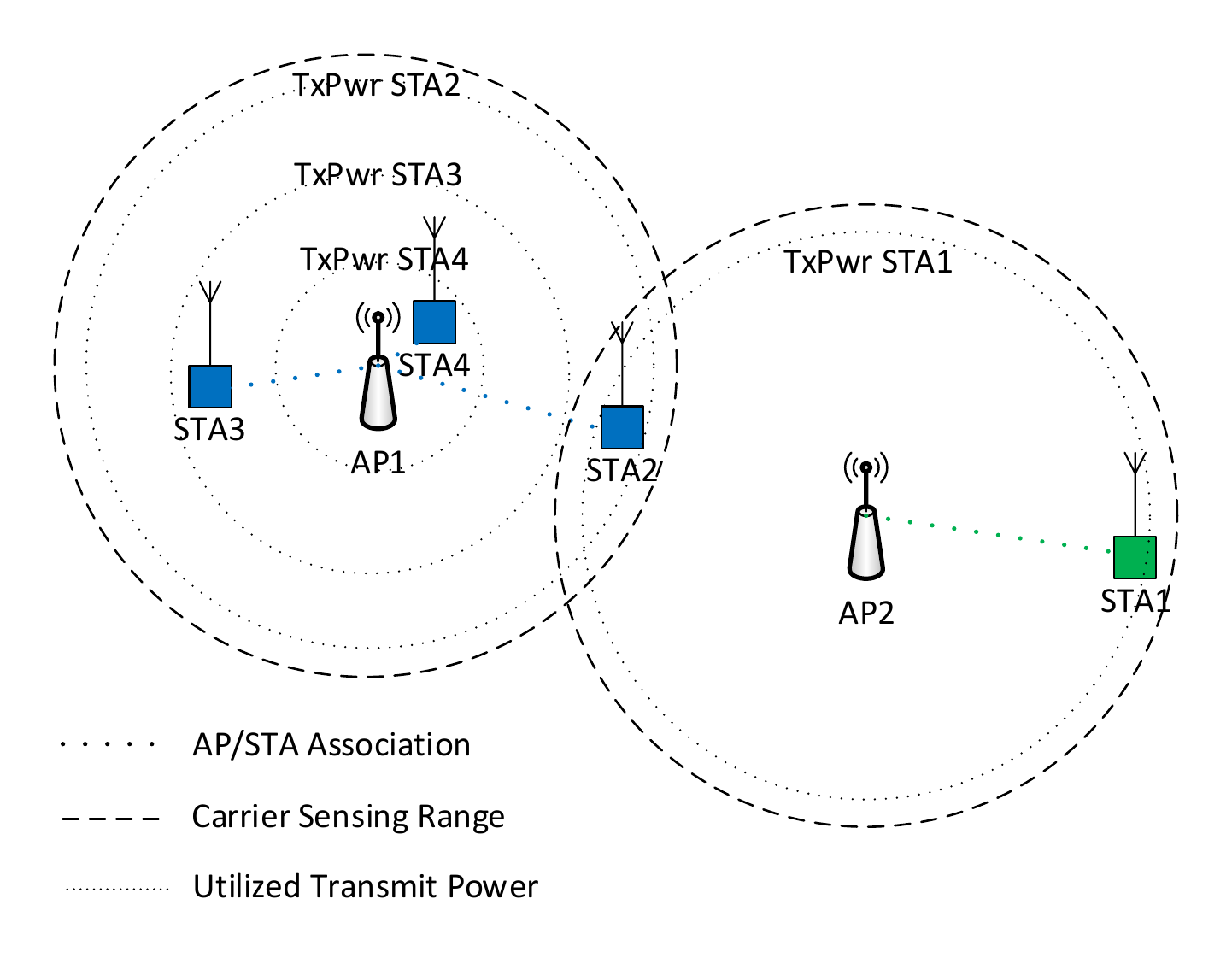}
	\end{center}
	\vspace{-10pt}
	\caption{Example 3 - Frame collisions only occur if the link AP1-STA2 and AP2-STA1 are served by the APs simultaneously, while the links AP1-STA3, AP1-STA4 and AP2-STA1 can be served simultaneously without creating any colissions (only considering downlink traffic). In addition, AP2 and STA2 may be in the exposed node scenario if uplink traffic of STAs would be considered and STA2 uses compareable transmission power as AP2.}
	\label{fig:link-tpc}
\end{figure}

Finally, a similar situation appears in networks that perform \textbf{transmit power} control on a per-link or per-packet basis, e.g. as suggested by \cite{Jung2002,Akella2007, Subbarao99dynamicpower-conscious,yeh-2003,huehn2012practical, huehn2010joint,huhn2013measurement,lowcostmesh2012}. Fig.~\ref{fig:link-tpc} illustrates a scenario where per-link power control leads to hidden node problem. Again, it can be resolved in an efficient way by assignment of exclusive time slots on a per link basis (Fig.~\ref{fig:link-tdmab}).

\medskip

In this technical report, we present hMAC, a software only solution that enables the assignment of TDMA time slots on a per link basis while still using the underlying CSMA/CA mechanism as specified in the IEEE 802.11 standard. The proposed hMAC enables to resolve efficiently hidden node problems while not unnecessarily preventing spatial reuse which is the case with classical TDMA schemes known from literature. Moreover, by keeping CSMA/CA the hMAC is still being compatible to the 802.11 standard.

We tested our hMAC prototype in a small 802.11 testbed in a hidden node scenario and compared it to standard 802.11 DCF and classical TDMA. Finally, we provide our implementation as open-source under \textbf{https://github.com/szehl/ath9k-hmac/}.




\section{Related Work}\label{sec:related_work}

The solutions for enabling TDMA on top of COTS 802.11 hardware can be classified according to solutions implemented in wireless network device drivers and those implemented above utilizing OS's traffic control mechanisms to schedule outgoing network traffic.

The most recent work on TDMA on top of COTS hardware is presented in OpenTDMF~\cite{yang2015enabling}. OpenTDMF enables per flow scheduling using a centralized controller and software agents on the APs. In contrast to all other approaches, OpenTDMF provides additionally a possibility to not only apply TDMA to the downlink traffic but rather also to enable TDMA for the uplink traffic. Both downlink and uplink TDMA is enabled by exploiting hardware registers that have been originally designated to enable the operation of the hybrid coordination function (HCF). Of course, this approach requires extensive modification on both, the clients and the APs and is therefore not compatible with legacy 802.11 devices.

The MadMAC approach~\cite{sharma2006madmac}, was built on top of the Linux MadWiFi wireless driver and is fully implemented in kernel-space. As arbitrary slot sizes can be configured, MadMAC uses guard intervals to ensure that packets are not truncated during transmission. In FreeMAC~\cite{sharma2008freemac}, the authors in addition to TDMA also provide the possibility to do channel switching between different slots. To ensure accurate scheduling, FreeMAC utilizes periodic hardware interrupts, generated by the Wi-Fi device, which are usually used for the beacon interval. Moreover, to bypass the random backoff delay, FreeMAC sets the values of CWmin and CWmax to zero. Emmelmann et al.~\cite{emmelmann2008system} aim at providing seamless handover by using a slotted TDMA MAC for downlink traffic. The TDMA MAC is implemented within the firmware of COTS hardware of the vendor IHP with drivers for Linux based systems.

Another solution is presented in \cite{djukic2009soft}, in which scheduling and controlling is done in user-space, while the module in kernel-space is responsible for finding an appropriate slot for the waiting transmissions. A full protocol is defined which enables microsecond time synchronization between all participating nodes. As the slotting is bound to the transmission time of OFDM symbols at different rates, no guard intervals are needed.

Pseudo TDMA (pTDMA) was introduced by Lee et al.~\cite{Lee-2014} in the context of resource slicing in Software-defined Networking (SDN). pTDMA uses Linux traffic control, i.e. a modified Qdisc, to control the time at which egress packets are transmitted. In order to control uplink traffic the pTDMA software need to be installed on the client devices as well. As pTDMA is implemented on a high layer the time granularity for time slots is course, i.e. 10\,ms.

\section{Background Information}\label{sec:wifiarch}

This section gives a brief overview of the relevant aspects of the Wi-Fi stack implementation inside the Linux kernel specifically the ATH9K wireless driver module and its queuing functionality is discussed. Moreover, a brief overview of the 802.11 power saving functionality is given.

\subsection{ATH9K SoftMAC Driver}

The ATH9k driver is part of the so-called SoftMAC architecture of the Linux Wi-Fi stack. Originally, Wi-Fi MACs were completely implemented as FullMAC devices, in which all MAC layer functions are controlled by the individual hardware or firmware of the device. In contrast to FullMAC devices in a SoftMAC device the whole 802.11 frame management has to be done by a software module running on the host system. 
 
\begin{figure}[!ht]
	\begin{center}
		\includegraphics[width=0.4\linewidth]{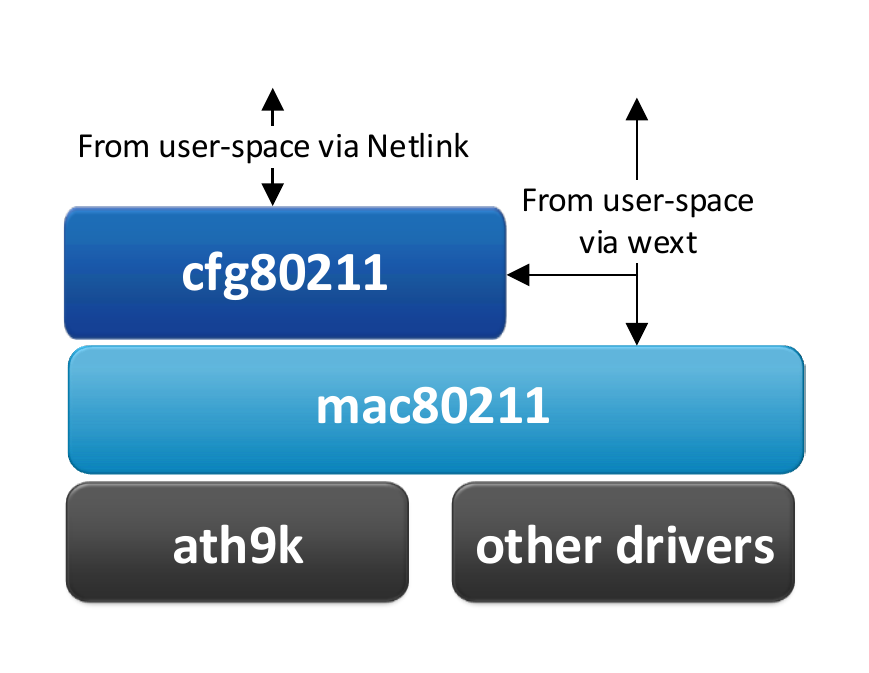}
	\end{center}
	\vspace{-10pt}
	\caption{Mac80211 Structure.}
	\label{fig:softmac}
\end{figure}

The Linux Wi-Fi SoftMAC module consists of general hardware independent modules that are shared by all underlying Wi-Fi driver modules (cfg80211 and mac80211) and additional hardware dependent modules such as the ATH9K driver module, cf. Fig.~\ref{fig:softmac}. 

Communication from user-space to the kernel-space is enabled by the use of Netlink sockets and a set of predefined commands called NL80211. All commands are received by cfg80211, which exists as layer between user-space and mac80211. Downwards, the mac80211  module is communicating with the specific hardware driver, e.g. ATH9K for Atheros cards or iwlwifi for Intel cards. All the kernel internal communication between cfg80211 and mac80211 and between mac80211 and the hardware driver is realized by callback functions.

\subsection{ATH9k Frame Queuing}

In order to enable aggregation, block acknowledgement and power saving functionality, the ATH9K driver is maintaining software queues for each link and each traffic identifier (TID). TID values between 0 – 7 are priorities that are mapped to the corresponding 802.11e traffic class hardware queues (background, best effort, voice and video). TID values between 8 – 16 are used to implement the 802.11e traffic specification (TSPEC) functionality.

\subsection{IEEE 802.11 Power Saving Functionality}

The standard 802.11 power saving mechanism (PSM) enables client STAs to sleep when it has no frames to send and no frames destined to it are waiting to be received. In this doze state, the STA is able to save energy, but is not able to transmit or receiver any frames. If there are frames destined to a sleeping STA operating in infrastructure mode, the AP buffers these frames till the STA indicates that it is awake and ready to receive them. In the opposite direction, a STA is always able to transmit frames destined to the AP.

To enable this functionality, the periodic beacon frames of the AP include a Traffic Indication Map (TIM) which indicates if the AP has buffered frames for each associated and PSM enabled client STA. The sleeping STAs wake up periodically to evaluate the beacon frames and the TIM. If there are frames buffered for a STA, it indicates its readiness to receive by sending a PS-POLL frame to the AP. The AP then dequeues all frames destined to this STA. To let the STA know when it is able to sleep again, a More Bit (set to one if more frames are waiting) within each frame is used.

\section{hMAC -- Design Principles}\label{sec:hmac_design}

%
%
\subsection{System Model}

\begin{figure}[!ht]
	\begin{center}
		\includegraphics[width=0.7\linewidth]{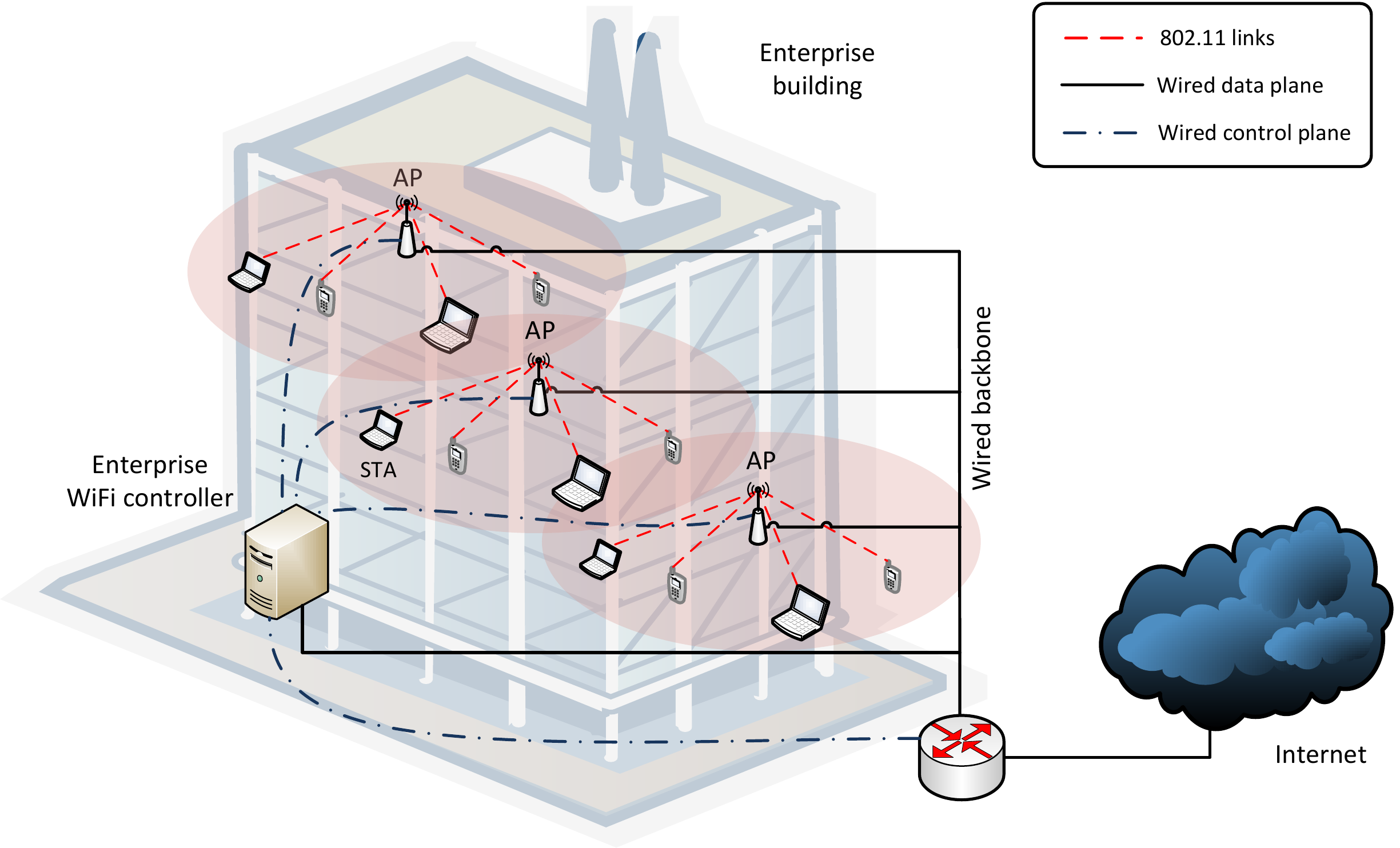}
	\end{center}
	\vspace{-10pt}
	\caption{Enterprise Wi-Fi network.}
	\label{fig:enterprise_wifi}
\end{figure}

Infrastructure Wi-Fi deployments are the topology of choice in most enterprise settings (Fig.~\ref{fig:enterprise_wifi}). These deployments usually consist of several basic service sets (BSS), consisting of one AP and several client STAs, that form an extended service set (ESS). Moreover, we assume that all APs comprising the ESS are under control of a single entity.

In advance, all APs of every BSS within the ESS are assumed to have in addition to their wireless interface a wired interface that can be used as a side-channel to achieve fine grained time synchronization e.g. using the precise time synchronization (PTP) protocol.

%
%
\subsection{Design Principles}

The goal of hMAC is to enable a simple software only solution that provides the possibility to pause and unpause downlink traffic on single APs on a per link basis for interference management reasons, i.e. efficiently solving hidden nodes scenarios, cf. Fig.~\ref{fig:hidden-node}, data/ACK collision scenarios, cf. Fig.~\ref{fig:no-hidden-node}, or hidden node scenarios in per link power controlled deployments, cf. Fig.~\ref{fig:link-tpc}.

Moreover, to provide interoperability with legacy STAs the standard CSMA/CA must still be active, which is contrary to most of the related work, cf. Sec. \ref{sec:related_work}. For this reason, hMAC utilizes the existing ATH9K driver power saving implementation to pause and unpause the traffic per link. As the power saving functionality is running on the host, no modifications to the registers of the Wi-Fi device need to be done which preserves the standard MAC functionalities on the device such as CSMA/CA. For this reason, the hMAC does not introduce additional unfairness and is fully compliant with the IEEE 802.11 standard.

%
%
\subsection{Application Programming Interfaces}

\begin{figure}[!ht]
	\begin{center}
		\includegraphics[width=0.9\linewidth]{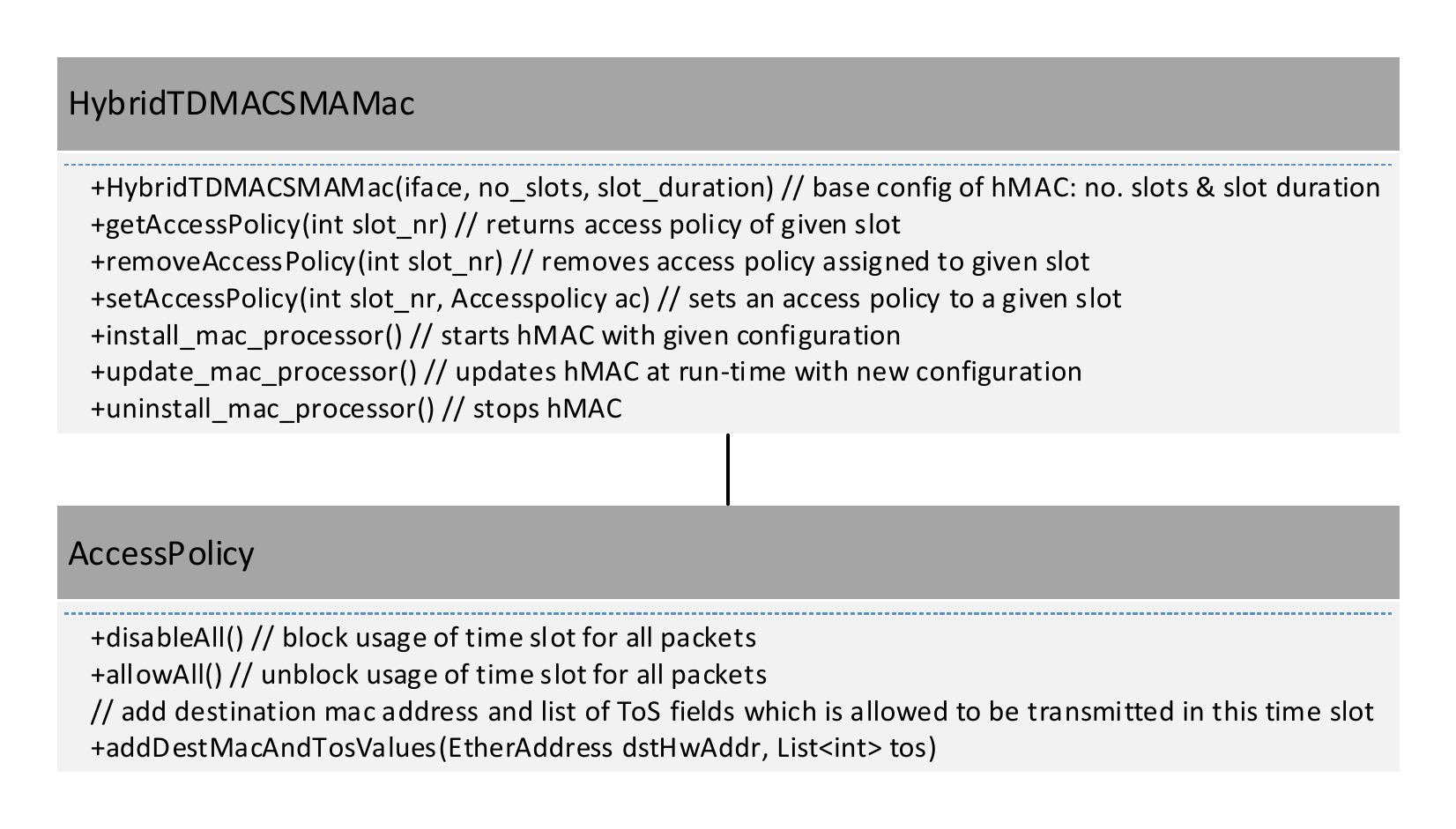}
	\end{center}
	\vspace{-10pt}
	\caption{hMAC Python API (UML class diagram).}
	\label{fig:uml}
\end{figure}

The hMAC exposes an application programming interface (API) to abstract the underlying details and enable easy control of the underlying functionality depending on the application requirement. This API is described as class diagram in Fig.~\ref{fig:uml}. In order to configure hMAC the user has to create an instance of the \textit{HybridTDMACSMAMac} class. In the constructor the general parameters of the hMAC are defined such as the total number of slots in superframe and their duration. Further the user can define the access policy for each TDMA slot in the superframe by creating an instance of the \textit{AccessPolicy} class and set it using \textit{setAccessPolicy()} function. Note, the access policy is defined as a set of next hop MAC addresses and type of service (ToS) values. This allows controlling the medium access within a time slot on a per-link and per-ToS basis. Note, the Python library internally maps the ToS values to the TIDs.

\section{hMAC -- Implementation Details}\label{sec:hmac_implementation}

We implemented hMAC by adding two additional functions to the ATH9K wireless driver. The first function can be used to pause distinct software queues of the ATH9K driver which are identified by the receiver MAC address and traffic identifier (TID). The second function then enables to continue distinct queues. For calling the functions we utilized the Netlink protocol and extended therefore the standard functionalities of nl80211 to enable the call of these two functions from Linux user-space.

\begin{figure}[!ht]
   \begin{center}
       \includegraphics[width=0.6\linewidth]{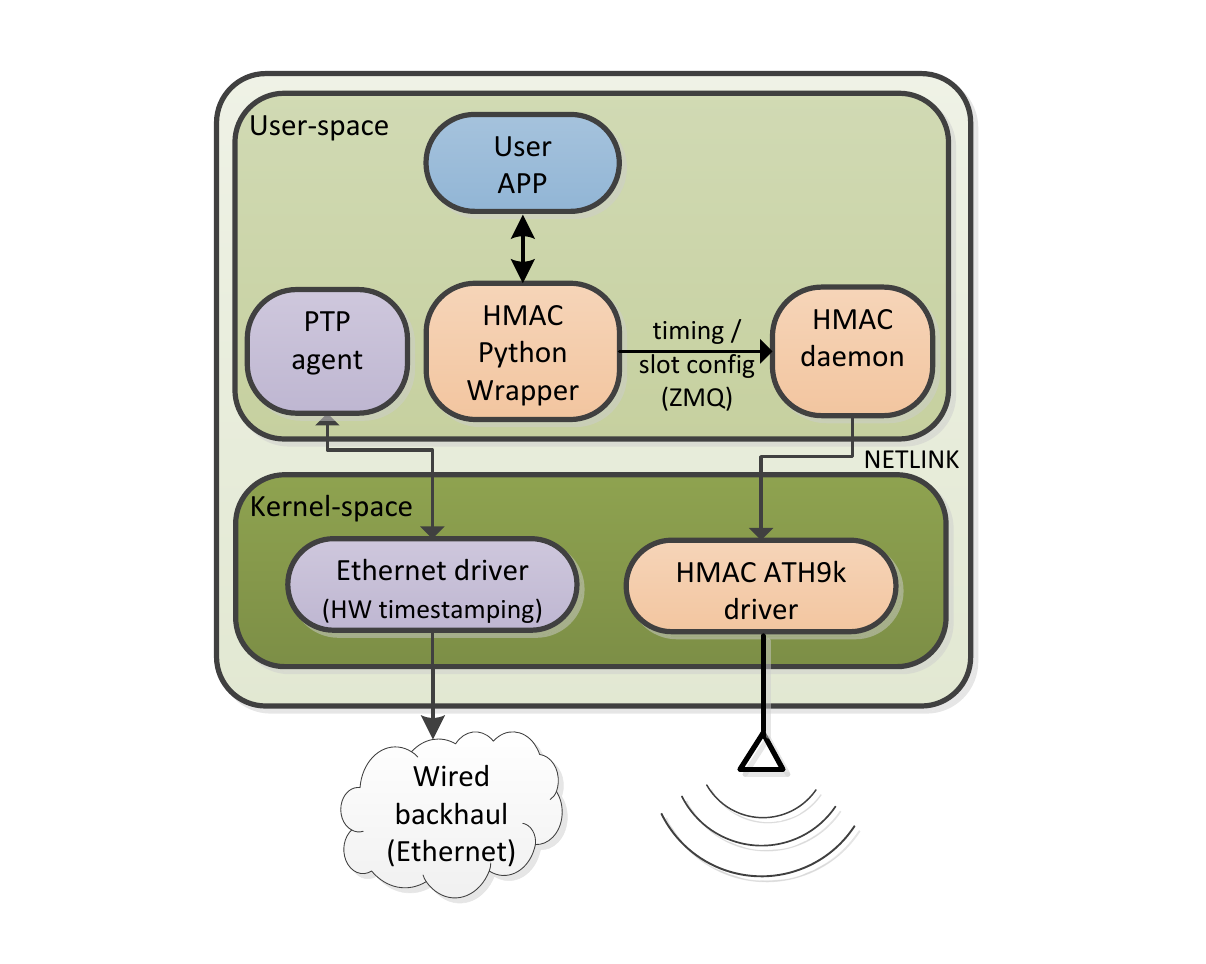}
   \end{center}
    \vspace{-10pt}
   \caption{hMAC implementation.}
   \label{fig:hmac_impl}
\end{figure}

Within user-space, a small C++ daemon equipped with a Netlink socket takes care of the slot scheduling. All participating APs are time synchronized using the PTP protocol which is running via the wired Ethernet interface. We run a central entity which is acting as PTP master. In order to align the beginning of the time slots on different nodes, we simply use the beginning tick of a new second as reference and adjust the slots after every superframe to it. In every slot, the daemon is sending the current slot configuration directly via Netlink to the ATH9K driver.

In addition to the Netlink socket, the daemon is also equipped with a ZeroMQ socket that allows easily controlling the daemon at runtime via the API shown in Fig.~\ref{fig:uml}. This API is implemented as Python object and can be easily integrated in any Python project, cf. Fig. \ref{fig:hmac_impl}.

\medskip

The source code of our prototype is available at \textbf{https://github.com/szehl/ath9k-hmac/}. On a standard Linux Ubuntu system it takes just three steps to install hMAC. Moreover, we provide installation instructions for other Linux distributions and Linux kernels.

\section{Example Usage of the hMAC API}\label{sec:hmac_examples}

The example configuration in Listing~\ref{hmac_hello_python} shows how to use the hMAC Python API. Here, a simple schedule with a 10 time slot superframe and a single slot duration of 20\,ms on interface \textit{wlan0} is created. Afterwards the first four slots are configured with type of service 0, which means best effort, and the destination MAC address (link identifier). Moreover, the access policy is applied by using the \textit{setAccessPolicy()} function. Besides the four best effort slots, all other slots are blocked, which means no traffic is allowed on the other slots. If this configuration would be used to mitigate a hidden node problem, we would simply change the best effort against the blocked slots and execute this setting on the other node.

\lstset{language=Python, 
        basicstyle=\ttfamily\scriptsize, 
        keywordstyle=\color{deepblue},
        commentstyle=\color{comments},
        stringstyle=\color{deepgreen},
				emphstyle=\ttb\color{deepred},    
        showstringspaces=false,
        procnamekeys={def,class}}
 
\begin{lstlisting}[caption=Example setting up hMAC using Python., label=hmac_hello_python]
#!/usr/bin/env python
import time
from hmac_python_wrapper import HybridTDMACSMAMac, AccessPolicy

if __name__ == "__main__":

    # configuration of hybrid MAC
    dstHWAddr = "34:13:e8:24:77:be" # STA destination MAC address
    total_slots = 10 # number of slots in superframe
    slot_duration = 20000 # each slot is 20 ms
    iface = 'wlan0' # Wi-Fi interface

    # create new MAC for local node
    mac = HybridTDMACSMAMac(log, iface, total_slots, slot_duration)

    be_slots = [1,2,3,4]
    # assign access policies to each slot in superframe
    for slot_nr in range(total_slots):
        if slot_nr in be_slots:
            # those are slots for best effort traffic towards our STA
            acBE = AccessPolicy()
            acBE.addDestMacAndTosValues(dstHWAddr, 0)
            mac.setAccessPolicy(slot_nr, acBE)
        else:
            # those are guard slots
            acGuard = AccessPolicy()
            acGuard.disableAll() # guard slot
            mac.setAccessPolicy(slot_nr, acGuard)

    # starting hMAC with above configuration
    mac.install_mac_processor()
		
    time.sleep(20)
		
    # Stopping hMAC
    mac.uninstall_mac_processor()
\end{lstlisting}

\section{Evaluation}\label{sec:evaluation}

The prototypical hMAC implementation was analyzed by means of experiments in a small Wi-Fi testbed.

%
%
%
\subsection{Methodology}

We set-up a hidden node scenario as described by Example 1 in the introduction section, cf. Sec.\ref{sec:introduction}, and depicted in Fig~\ref{fig:hidden-node}. Here the two APs, AP1 and AP2, are out of carrier sensing range, i.e. cannot sense each other. Hence, when AP1 sends DL frames to STA2 while AP2 sends DL frames to STA1, AP2's frames will collide at STA2, resulting in decreased throughput on link AP1-STA2. As already mentioned, a classical TDMA would separate the DL traffic of AP1 and AP2 in time domain by applying e.g. the schedule shown in Fig.~\ref{fig:classical_tdma}. However, as such a scheme is inefficient for APs serving multiple client stations, cf. Sec.\ref{sec:introduction}, we utilize our hMAC with per-link TDMA configuration using the schedule shown in Fig.~\ref{fig:per_link_tdma_sched}. As baseline, the unmodified ATH9K driver using standard 802.11 DCF is used. Then to show the current state of the art, classical TDMA with the schedule shown in Fig.~\ref{fig:classical_tdma} is applied. Finally, our proposed hMAC with per-link TDMA scheduling is used. In case of TDMA a slot size of 20\,ms was used. Moreover, the total number of slots in a superframe was set to 10 and 4 slots were assigned to each AP, while 2 slots serve as guard slots. 

The experiment is conducted using off-the-shelf (COTS) Atheros AR9280 chipsets (802.11n) connected via PCI-Express and executed on x86 machines with Ubuntu 14.04 LTS. The APs were time synchronized using the Precision Time Protocol (PTP). In all experiments we measured the simultaneous DL TCP throughput on the links AP1 to STA2, AP1 to STA3 and AP2 to STA1 using the iperf tool~\cite{iperf-2013}.

\subsection{Results}

\begin{figure}[!ht]
   \begin{center}
       \includegraphics[width=0.9\linewidth]{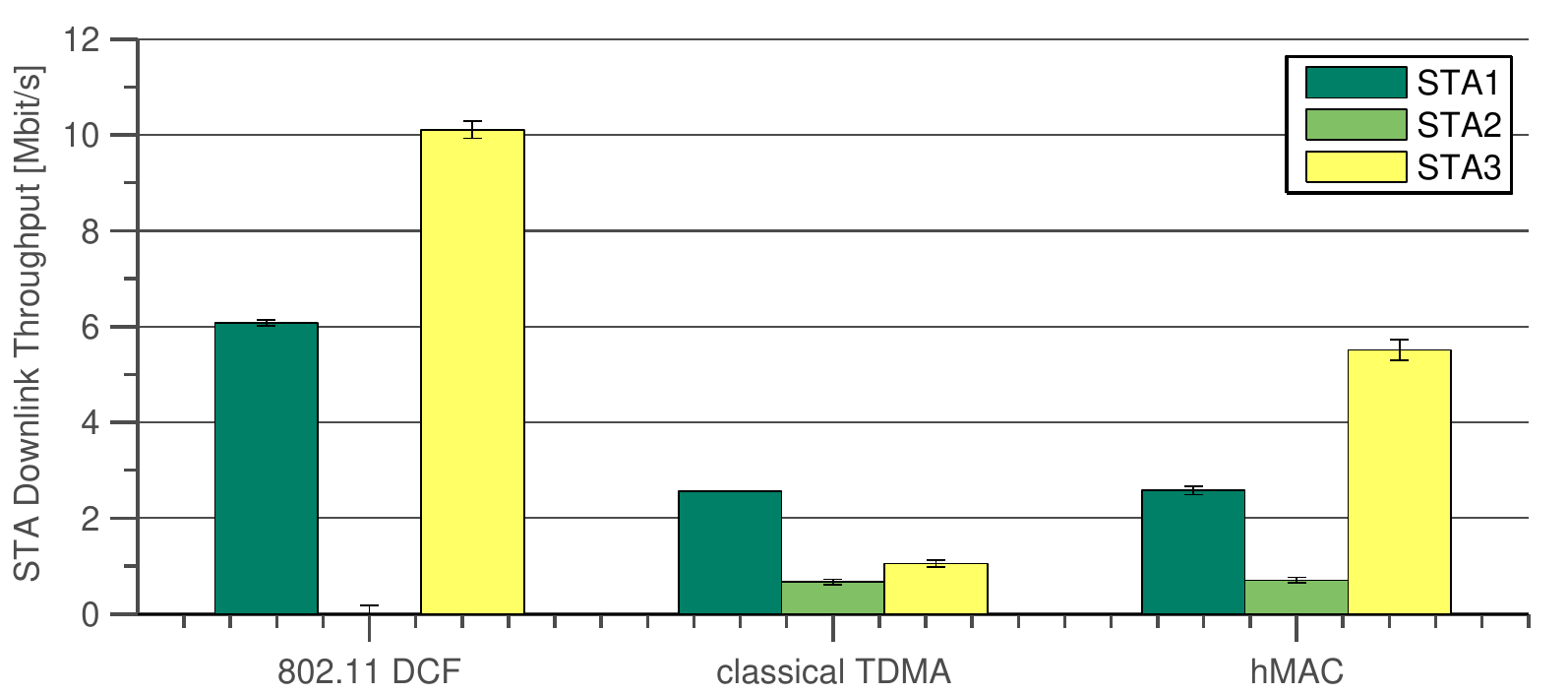}
   \end{center}
    \vspace{-10pt}
   \caption{STA DL throughput for the hidden node scenario depicted in Fig~\ref{fig:hidden-node} for 802.11 DCF, classical TDMA and proposed hMAC (error bars show standard error).}
   \label{fig:hmac_hidden_node_exp}
\end{figure}

Fig.~\ref{fig:hmac_hidden_node_exp} shows the results of the experiment. As expected, due to the hidden node scenario, the frames of the link AP2 - STA1 collide with the frames of the link AP1 - STA2, which results in near zero throughput towards SAT2 (only malformed frames were received), cf. Fig.~\ref{fig:hmac_hidden_node_exp}, when using standard 802.11 DCF.

When classical TDMA scheme is applied, the hidden node problem is solved and the link AP1-STA2 is able to achieve a DL throughput of about 1\,Mbit/s, cf.~Fig. \ref{fig:hmac_hidden_node_exp}. 
However such a schedule is inefficient, i.e. it created a disadvantage for the link AP1-STA3 which was only able to transmit at the quarter of the available air-time (half because of TDMA and quarter because AP1 is only able to server both links round robin) and therefore was only able to achieve a TCP DL throughput of about 1.2\,Mbit/s.

This drawback is not present if the hMAC with the per-link schedule as depicted in Fig.~\ref{fig:per_link_tdma_sched} is applied. As AP1 is now able to serve STA3 within the full duty-cycle, a TCP DL throughput of about 5.5\,Mbit/s is achieved, while the throughput for the links AP1-STA2 and AP2-STA1 stays constant. Overall, using the hMAC approach for this distinct scenario results in a doubling of the total (sum) throughput, 4.2 vs. 8.8\,Mbit/s, as compared to classical TDMA.

\medskip

\section{Known Limitations and Possible Enhancements}\label{sec:discussion}

The hMAC scheduler residing in user-space controls at the beginning of each TDMA time slot the pausing or unpausing of the software queues of the ATH9k driver using the Netlink protocol. Such a realization has its pros and cons. On the one hand, the modifications within the driver residing in kernel-space are as small as possible and the development of scheduling programs within user-space is easier than in kernel-space. On the other hand, the message transfer between user-space and kernel-space is non-deterministic and if the host is under high load, accurate timing for short slot intervals cannot be guaranteed. During evaluation of the hMAC implementation, cases occurred in which due to a high amount of interrupts, the ATH9K driver was not able to handle all incoming hMAC Netlink messages in time. As every wake up and sleep command is first generated by the scheduler in user-space and then sent via Netlink to the ATH9K driver in kernel-space this results in too short or too long slots (because of queued Netlink messages). This limitation can be bypassed if large slot sizes are used, e.g. $>10$\,ms and guard slots are used, e.g. as suggested in~\cite{sharma2006madmac}. Nevertheless, if accurate timing in combination with small slot sizes $<10$\,ms is needed, the time scheduling of the slots should be ported to kernel-space and also bound to hardware interrupts, e.g. the beacon hardware interrupts as suggested in~\cite{sharma2008freemac}.

\section{Conclusions \& Future Work}\label{sec:conclusions}

In this technical report we presented the first preliminary results of our current work-in-progress on a hybrid TDMA/CSMA/CA Wi-Fi MAC implementation called hMAC. The software-only solution exploits the existing 802.11 power saving functionality of the popular ATH9K SoftMAC driver to enable per link and per traffic identifier pausing and unpausing of the software queues. As this approach is done above the actual Wi-Fi hardware, it does not change any part of the standard 802.11 medium access. This ensures compatibility and fairness between legacy non hMAC STAs. Moreover, we presented first preliminary experimental testbed results of the performance of hMAC in comparison to classical per-node TDMA and 802.11 DCF. Finally, we published the hMAC implementation as open source to enable collaboration with other researchers.

For our future work we plan to use hMAC on top of the ResFi framework~\cite{resfiwowmom2016} for interference management in residential Wi-Fi networks where the Internet-scale latencies pose a significant challenge as compared to Enterprise networks. Moreover, with NxWLAN~\cite{Gawlowicz16nxwlan} we introduced an architecture which enables mutual residential Wi-Fi sharing. NxWLAN makes use of virtualization techniques. We argue that radio resources in the AP should be dynamically sliced based on traffic among visiting client stations depending on the neighbors policy which could be easily achieved using hMAC. Finally, having the possibility to control the software queues of distinct 802.11 links, enables to improve our existing BigAP~\cite{bigap} soft-handover scheme in terms of frame loses during the handover operation, e.g. in order to avoid packet losses the packet queue on target AP is paused just before handover operation and continued after handover is completed, cf. \cite{bigap, bigap2}.

\section*{Acknowledgment}
This work has been supported by the European Union’s Horizon 2020 research and innovation programme under grant agreement No. 645274 (WiSHFUL project).

\bibliography{biblio}

\end{document}